# Exceptional points and lasing self-termination in photonic molecules

R. El-Ganainy[1,*], M. Khajavikhan[2], and Li Ge[3,4,†]

[1]*Department of Physics, Michigan Technological University, Houghton, Michigan, 49931, USA*

[2]*College of Optics/CREOL, University of Central Florida, Orlando, Florida 32816, USA*

[3]*Department of Engineering Science and Physics, College of Staten Island, CUNY, Staten Island, NY 10314, USA*

[4]*The Graduate Center, CUNY, New York, NY 10016, USA*

[*]ganainy@mtu.edu

[†]li.ge@csi.cuny.edu

**Abstract**

We investigate the rich physics of photonic molecule lasers using a non-Hermitian dimer model. We show that several interesting features, predicted recently using a rigorous steady state ab-initio laser theory (SALT), can be captured by this toy model. In particular, we demonstrate the central role played by exceptional points in both pump-selective lasing and laser self-terminations phenomena. Due to its transparent mathematical structure, our model provides a lucid understanding for how different physical parameters (optical loss, modal coupling between microcavities and pump profiles) affect the lasing action. Interestingly, our analysis also confirms that, for frequency mismatched cavities, operation in the proximity of exceptional points (without actually crossing the square root singularities) can still lead to laser self-termination. We confirm this latter prediction for two coupled slab cavities using scattering matrix analysis and SALT technique. In addition, we employ our model to investigate the pump-controlled lasing action and we show that emission patterns are governed by the locations of exceptional points in the gain parameter space. Finally we extend these results to multi-cavity photonic molecules, where we found the existence of higher-order EPs and pump-induced localization.

PACS#  42.55.Sa, 42.55.Ah, 42.60.Da

# 1. Introduction

Optical microcavities and their potential applications in science and industry has been a subject of intense investigations in the past decades [1,2]. One particular interest is building low threshold compact laser systems, as demonstrated in different physical platforms such photonic crystals [3,4], microdisk resonators [5,6], and more recently in plasmonic nanocavities [7]. These works were later extended to investigate lasing action in more than one microcavity, evanescently coupled to form photonic molecules [8,9]. In principle, the performance of these laser systems can be studied by numerical simulation of the Maxwell's Bloch (MB) equations [10,11]. This, however, requires heavy computational resources and thus other techniques were developed to achieve this task. For instance, high quality factor cavities can be effectively studied using the cold cavity method, i.e. treating each quasi-bound mode as a lasing mode [12]. Recently, a Steady-state Ab-initio Laser Theory (SALT) was developed and shown to provide accurate results for both regimes of high and low quality factor cavities, which captures the openness of the cavity exactly and treats multimode interactions to infinite order [13,14]. When compared with time-dependent simulations of the MB equations, SALT is up to $10^3$ times faster for the same laser cavity [15,16]. Equipped with this numerical tool, more investigations of microlasers have been recently performed and several new features of these systems have been identified [17,18]. For example, it was shown that asymmetric optical pumping can be used to manipulate the lasing threshold of different modes in random lasers [17,19,20]. Moreover, careful simulation of laser action in two-cavity photonic molecules revealed the interesting property of self-terminations of lasing modes [18]. In this process, lasing action is achieved by pumping only one cavity and shuts down when this gain level is kept constant while gradually switching on the pump in the second resonator. Similar behavior was subsequently shown to occur in different physical systems such as electronic circuits [21]. We note that laser self-termination is a counter-intuitive effect given the fact that the overall pumping of the system has increased. The crucial role played by the exceptional points (EPs) of the non-Hermitian system in explaining this behavior was highlighted in Ref. 18, which indicates that it is a linear effect. This analysis provides a numerical evidence for these results as well as a beautiful physical insight into laser self-termination process. However, it relies on the finding of the EPs numerically instead of predicting their whereabouts in terms of simple system parameters.

Here in this work, we employ a toy model based on non-Hermitian dimers in order to gain more insight into photonic molecule lasers and their EPs. While our model cannot in principle substitute for the more rigorous SALT analysis, it captures all the effects discussed in Ref. 18 transparently, in terms of cavity detuning and decays rates, the asymmetric gain profile, and the inter-cavity coupling $J$. More specifically, we show that the formation of the EP is due to a square root singularity of the difference between $J^2$ and the asymmetric gain parameter $\Delta\gamma^2 = (\gamma_a - \gamma_b)^2/4$, with $\gamma_{a,b}$ being the gain coefficients in the two coupled cavities. The self-termination behavior can then be easily analyzed in the two-dimensional pump space $(\gamma_a, \gamma_b)$, in which there is a continuous EP curve defined by $\Delta\gamma = J$. Along any monotonically increasing pump trajectory that leads to self-termination, *two* EPs are encountered instead of one; the self-termination occurs after passing the first EP (when $\Delta\gamma$ increases with the pump in the first cavity) and before reaching the second EP (when $\Delta\gamma$ decreases with the increasing pump in the second cavity). Our analysis also reveals that, for frequency mismatched cavities, operation in the proximity of exceptional points (without actually crossing the square root singularities) can still lead to laser self-termination. The intuitive understanding for the interplay between different physical parameters enables us to probe more complicated systems, exemplified using a four-cavity array where a fourth-order EP can be realized and pump-induced localization occurs in succession to the laser self-termination.

## 2. Pump-controlled laser emission and self-termination

We consider a non-Hermitian optical dimer made of two coupled cavities with loss and subject to asymmetric pumping:

$$i\frac{da}{dt} = \omega_a a + i(\gamma_a - \kappa_a)a + Jb, \qquad (1.a)$$

$$i\frac{db}{dt} = \omega_b b + i(\gamma_b - \kappa_b)b + Ja, \qquad (1.b)$$

In Eq. (1), $a$ and $b$ are the field amplitudes of these two lasing modes and $\omega_{a,b}$ are their respective frequencies. In addition, $\gamma_{a,b}$ and $\kappa_{a,b}$ are the gain and loss associated of the two cavities and $J$ is the evanescent coupling coefficient. When $\omega_a = \omega_b = \omega_0$ (this condition does not necessarily imply identical cavities), Eq. (1) can be cast into the following form

$$i\frac{d}{dt}\begin{bmatrix}\tilde{a}\\ \tilde{b}\end{bmatrix} = H\begin{bmatrix}\tilde{a}\\ \tilde{b}\end{bmatrix}, \quad H = \begin{bmatrix} i(\Delta\gamma - \Delta\kappa) & J \\ J & -i(\Delta\gamma - \Delta\kappa)\end{bmatrix}, \quad (2)$$

via the transformation $[a(t)\ b(t)]^T = [\tilde{a}(t)\ \tilde{b}(t)]^T \exp(-i\omega_o t - \kappa_{avg} t + \gamma_{avg} t)$. Here $\Delta\kappa = (\kappa_a - \kappa_b)/2$ is the asymmetric loss parameter defined similarly to the asymmetric gain parameter $\Delta\gamma$, $\kappa_{avg} = (\kappa_a + \kappa_b)/2$ and $\gamma_{avg} = (\gamma_a + \gamma_b)/2$ is the average loss/gain parameter, respectively. Finally, the subscript $T$ denotes matrix transpose. We note that the effective Hamiltonian in Eq. (2) is Parity-Time symmetric [22-24], which implies the existence of EPs with a single-parameter tuning. Without loss of generality, we assume a zero asymmetric loss parameter, i.e. $\kappa_a = \kappa_b = \kappa$. This can be easily justified by offsetting the asymmetric gain parameter accordingly. The eigenvalues of Eq. (1) are then given by $\omega_\pm = \omega_o \pm \sqrt{J^2 - (\Delta\gamma)^2} + i(\gamma_{avg} - \kappa)$. The corresponding eigenfunctions are $V_\pm = \begin{bmatrix} 1 & \pm\sqrt{1 - (\Delta\gamma/J)^2} - i(\Delta\gamma/J)\end{bmatrix}^T$ up to a normalization constant, and we note that it is independent of $\gamma_{avg}$. From the expressions of $\omega_\pm$ and $V_\pm$, we immediately see that the EPs occurs when $\Delta\gamma = J$, at which $V_\pm = [1\ -i]^T$ and the intensity profile of the coupled modes is symmetric. In fact the latter property holds for all values of $\Delta\gamma \in [0, J]$, including the usually studied case where the pump is uniform ($\Delta\gamma = 0$). Thus we can define an intensity-symmetric "phase," which is broken once $\Delta\gamma > J$. In the broken-symmetry phase the relation $V_+^T V_- = const$ still holds, independent of $\Delta\gamma$. For $\Delta\gamma/J \to \infty$, $V_+ = [1\ 0]^T$ and $V_+ = [0\ 1]^T$, i.e. the inter-cavity coupling becomes negligible.

The laser threshold is reached when $\text{Im}[\omega_{\pm}]=0$, which leads to a self-sustained oscillation while for $\text{Im}[\omega_{\pm}]<0$, the supermodes decay with time. On the other hand when $\text{Im}[\omega_{\pm}]>0$, the system is above threshold and the amplitudes of the lasing modes are determined by the nonlinear terms not considered in Eq. (1). Since the laser self-termination is essentially a linear effect [18], the linear equation (1) is sufficient for our analysis here. We note that $\text{Im}[\omega_{+}]=\text{Im}[\omega_{-}]$ in the intensity-symmetric phase and $\text{Im}[\omega_{+}]>\text{Im}[\omega_{-}]$ in the symmetry broken phase. Consequently, the $V_{+}$ mode can always be treated as the first lasing mode in our linear formulism. As pointed out in [18], the strong nonlinear interaction between the two eigenfunctions $V_{+}$ and $V_{-}$ leads to a suppression of the latter. Thus we focus our attention only on the lasing $V_{+}$ supermode.

We first discuss the symmetry of the lasing mode at its threshold when pumping only cavity $a$. Depending on the relative strength of the overall loss of each cavity and the evanescent coupling between the two cavities, two distinct behaviors can take place. If $\kappa/J<1$, the lasing threshold at $\gamma_a=2\kappa$ precedes the exceptional point at $\gamma_a=2J$. As a result, the photonic molecule starts lasing in the intensity-symmetric phase. The situation is reversed if $\kappa/J>1$. The symmetric phase now only exists below the threshold, and the photonic molecule starts lasing in the symmetry broken phase. These two contrasting behaviors are depicted in Fig.1 for $J=1$ and $\kappa=0.5,1.5$, respectively. They highlights the fact that while lasing in cavities with high quality factors is largely predetermined by the passive modes or resonances, it can be controlled by non-uniform pumping in cavities having higher loss [17,19,20].

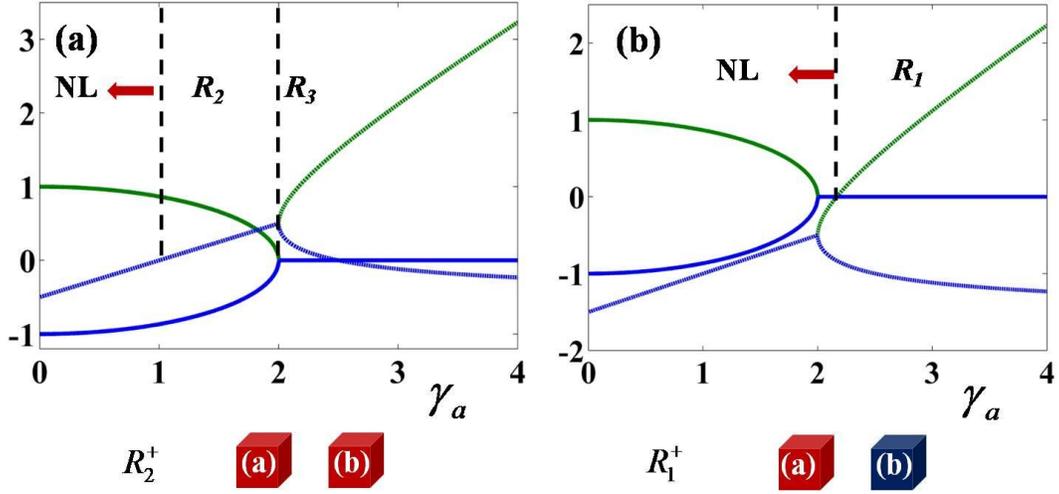

Fig.1. Pump-controlled laser action: real (curves starting at $\pm J$) and imaginary parts of $\omega_\pm - \omega_0$ (green/blue curves) as a function of the pump power $\gamma_a$ in cavity $a$ for a photonic molecule laser having inter-cavity coupling $J = 1$. In (a) $\kappa = 0.5$ and the $V_+$ mode starts lasing in the symmetric phase. Once above threshold the nonlinear effect is important, and the transition from $R_2$ to $R_3$ is smoothed or even suppressed. In (b) $\kappa = 1.5$ and the $V_+$ mode starts lasing in the asymmetric phase. The symmetry character of lasing at threshold in each regime is depicted schematically using colored cavities (cubes) where red/blue colors indicate high/low intensity and $R_j^+$ denotes the lasing of $V_+$ mode in region $R_j$ defined in the plots.

Next we consider pumping the two cavities of the photonic molecule sequentially when $\kappa/J > 1$. As the gain is increased along the trajectory $S_1$ (i.e. increasing $\gamma_a$ while keeping $\gamma_b = 0$), the asymmetric intensity mode is formed at the first crossing of the square root singularity (point $EP_1$) and reaches the lasing threshold at $L_1^+$. By further increasing the average gain along the path $S_2$ (i.e. increasing $\gamma_b$ from zero while keeping $\gamma_a$ at its maximum, $\gamma_{max}$), the system experiences lasing self termination (marked by the point $ST$) followed by a second exceptional point $EP_2$ before the symmetric eigenmode starts to lase as $L_2^+$. This behavior is depicted in Fig.2(a). The intensity distributions of the lasing modes at both points $L_{1,2}^+$ are shown schematically at Fig.2(b) where high/low intensities are indicated by the red/blue colors. We note that $EP_1$ and $EP_2$ are connected by a continuous line of EPs given by $\Delta\gamma = J$ as mentioned before.

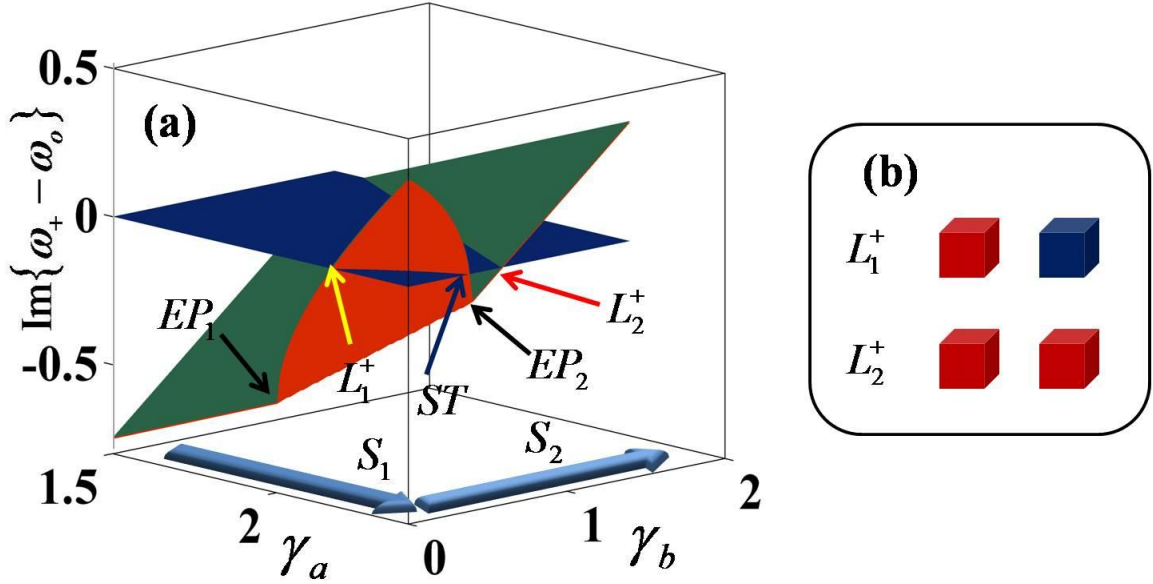

Fig.2. (Color online) Laser self termination: (a) depicts the imaginary part of the eigenmode $\omega_+$ for the parameters $J=1$ and $\kappa=1.5$. Self termination can be observed along the trajectory $S = S_1 \cup S_2$. As the gain is increased along the trajectory $S_1$ (only the portion starting from $\gamma_a = 1.5$ is shown), the system undergoes a phase transition at $\gamma_a = 2$ (denoted by $EP_1$), beyond which the intensity of the eigenmode becomes asymmetric (see (b); red/blue colors denote high/low intensities, respectively). At the point marked by $L_1^+$ where $\text{Im}\{\omega_+\}=0$, the asymmetric mode reaches the lasing threshold. Along the trajectory $S_2$, the system experiences three distinct stages: laser self termination ($ST$), second exceptional point $EP_2$, and lasing threshold at $L_2^+$ with a symmetric intensity profile (see (b)). For clarity, we used green/red colors for the top and bottom surface of the function $\text{Im}\{\omega_+ - \omega_o\}$ in Fig.2(a).

It is remarkable that such a simple model for photonic laser molecule can capture the laser self-termination characteristics as well as the general behavior of the lasing modes.

To further demonstrate the profound role of the exceptional points in lasing self-termination and the predictive power of our model, we consider a group of distinct pump trajectories $S = S_1 \cup S_2$ as in Fig.2 but with different values of $\gamma_{max}$. We note that on these trajectories the average gain parameter $\gamma_{avg}$ increases monotonically, but the asymmetric gain parameter $\Delta\gamma$ first increases from zero on $S_1$ and then decreases to zero on $S_2$. If $\gamma_{max} < 2J$ (i.e. $\max[\Delta\gamma] < J$), the photonic molecule is always in the intensity-symmetric phase, and there is only a single onset-threshold for the $V_+$ mode, similar to the conventional lasing behavior with a uniform pump profile [Fig.3(a)]. This lasing behavior holds even if $2J < \gamma_{max} < (\kappa + J^2/\kappa)$ and $\kappa > J$. In this case we encounter two EPs at $\gamma_a = 2J$, $\gamma_b = 0$ and $\gamma_a = \gamma_{max}$, $\gamma_b = \gamma_{max} - 2J$, respectively, similar to those in Fig. 2. Between them the system is in the symmetry-broken phase [Fig.3(b)]. Although $\text{Im}[\omega_+]$ peaks at $\Delta\gamma = \gamma_{max}$ in this phase, it still cannot reach the threshold until the system enters the symmetric phase again after passing $EP_2$. In both cases discussed above, the intensity of the $V_+$ mode increases monotonically after its onset (scenario I).

The occurrence of laser self-termination requires not just $\gamma_{max} > (\kappa + J^2/\kappa)$, i.e. the peak of $\text{Im}[\omega_+]$ in the symmetry-broken phase is above the threshold. More importantly, it also requires that $EP_2$ is below the threshold, which causes the self-termination [scenario II; Fig.3(c)]. This condition is achieved when $\gamma_{max} < (\kappa + J)$, which in turn requires $\kappa > J$ when compared with the lower bound of $\gamma_{max}$. The latter inequality highlights the necessity of having a relatively large loss for self-termination; Ref. 18 focused on the Terahertz laser for exactly the same reason (strong absorption per wavelength) but did not quantify this relation. If $\kappa < J$ or if $\kappa > J$ and $\gamma_{max} > (\kappa+J)$, $EP_2$ occurs above threshold [Fig. 3(d)] and the intensity of the $V_+$ mode experiences a drop in the vicinity of $EP_2$, without decreasing all the way to zero (scenario III).

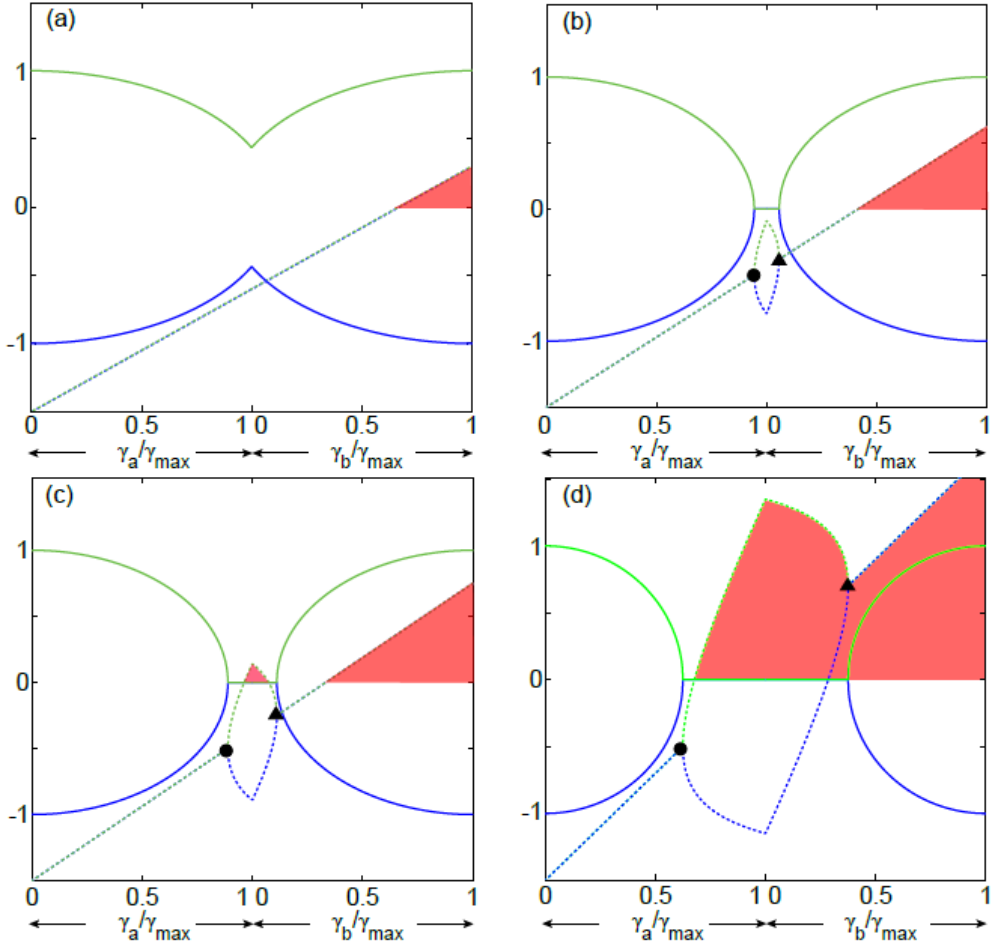

Fig.3. (Color online) Real (solid) and imaginary (dashed) part of $\omega_+ - \omega_0$ (green) and $\omega_- - \omega_0$ (blue) for $J=1$, $\kappa=1.5$ along different pump trajectories. Along each pump trajectory $\gamma_a$ is first increased to $\gamma_{max}$ and then fixed, followed by the increase of $\gamma_b$. $\gamma_{max} = 1.8 \in [0, 2J]$ in (a), $2.12 \in [2J, \kappa + J^2/\kappa]$ in (b), $2.25 \in [\kappa + J^2/\kappa, \kappa + J]$ in (c), and $3.2 \in [\kappa+J, \infty]$ in (d). The red regions show where the $V_+$ mode is lasing. The black dot and triangle points inidicate the $EP_1$ at $\gamma_a = 2J$, $\gamma_b = 0$ and $EP_2$ at $\gamma_a = \gamma_{max}$, $\gamma_b = \gamma_{max} - 2J$, respectively.

Evidently, our non-Hermitian dimer model capture all the scenarios (I, II, and III above) found in Ref. 18. This is a remarkable result given that this phenomenological model does not take into account all the processes involved in the lasing action, such as the line pulling of the cavity frequency towards the atomic transition frequency and the variation of the coupling between the two cavities along the pump trajectory. In order to understand this, we note that laser thresholds can be described by the poles of the corresponding linear scattering matrix, which are the eigenvalues $\omega_\pm$ in our simplified model. In particular, we consider the typical case of uniform pumping. Before adding any gain, the poles of a passive dimer in the absence of detuning are given by $\omega_\pm = \omega_0 \pm J - i\kappa$. Under uniform pumping conditions, these poles move vertically upward in the complex frequency plane in the direction of the real axis, as described in standard textbooks [10,11]. On the other hand, when the gain is added asymmetrically, the two poles move towards each other while approaching the real axis, as shown in Fig. 3 and similar to the finding of using ab-initio calculations of the scattering matrix in Ref. 20. Given this close relationship between $\omega_\pm$ and the scattering matrix, it is tempting to associate the EPs of non-Hermitian dimer Hamiltonian with those of the scattering matrix. However, this was shown to be not the case [25, 26].

Our model also predicts a few new features. For example, laser self-termination can still occur for *mismatched* cavities having different resonant frequencies, even though the EPs are not encountered exactly. Under the detuned condition $\Delta\omega = \omega_a - \omega_b$, the eigenvalues are given by $\omega_\pm = \omega_{avg} \pm \sqrt{J^2 - (\Delta\gamma)^2 + (\Delta\omega)^2 + 2i\Delta\omega\Delta\gamma} + i(\gamma_{avg} - \kappa)$, where $\omega_{avg} = (\omega_a + \omega_b)/2$. Evidently, in this case $\omega_\pm$ undergo avoided crossings in the complex plane, instead of coalescing at the two EPs. We note again that $\text{Im}[\omega_\pm]$ determines the thresholds of the lasing modes, and they behave similarly to the zero detuning case if $|\Delta\omega| \ll J$. This is illustrated in Fig.4, in which we revisit the cases considered in Fig.3 but now with a frequency detuning of 0.02. If the detuning is too strong, the avoided crossings are smoothed out and the lasing behaviors become similar to the case of uniform pumping. Thus it is precisely the detuning that hinders the possibility of self-termination in low-loss cavities.

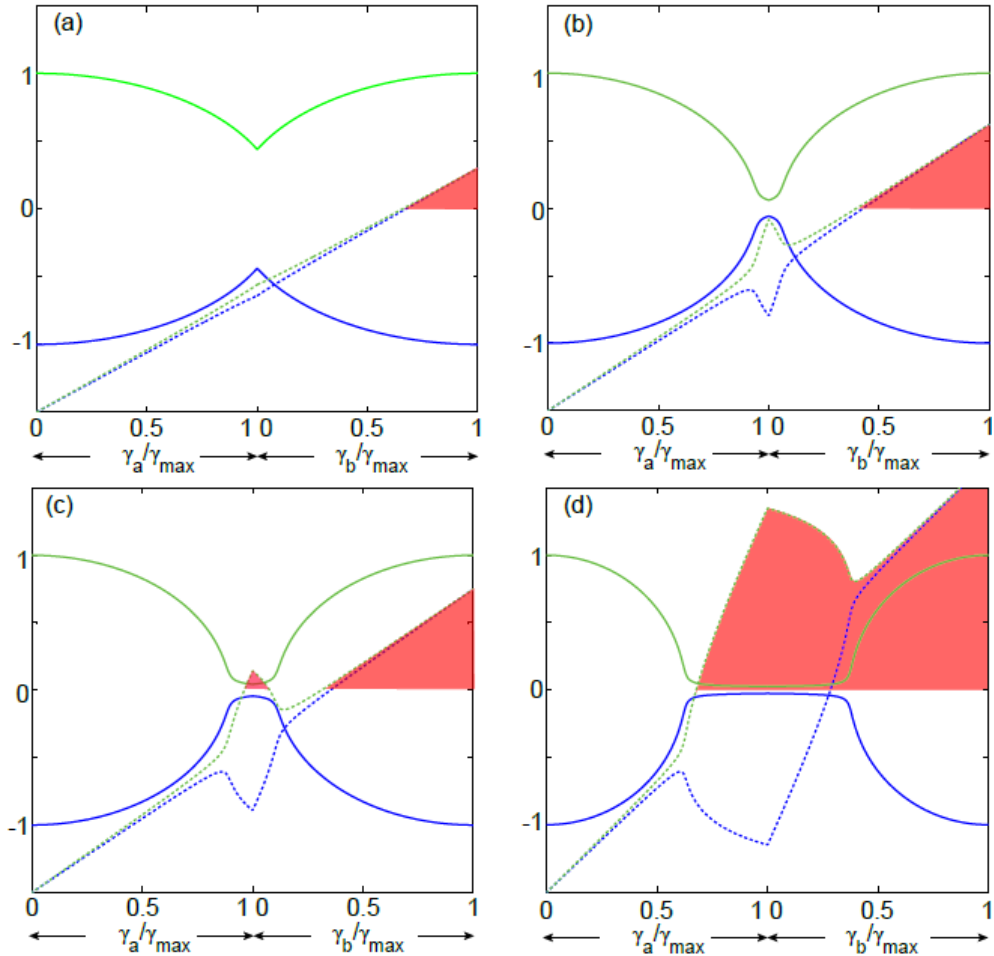

Fig.4 Real (solid) and imaginary (dashed) part of $\omega_+ - \omega_{avg}$ (green) and $\omega_- - \omega_{avg}$ (blue). The red regions show where the $V_+$ mode is lasing. Here $\Delta\omega = 0.02$ and the other parameters are the same as in Fig. 3.

To confirm the existence of laser self-termination in the presence of detuning, we calculate the threshold and laser intensity in a photonic molecule, consisting of two coupled one-dimensional cavities of length $L$ and refractive index $n = 3 + 0.007i$, with an air gap of width $L/10$ between them (inset; Fig. 5(a)). Using the approach in Ref. 20, we first calculate the poles of the scattering matrix, which are equivalent to $\omega_\pm$ in our simple model as mentioned previously. Moreover, we focus on the isolated modes (for each individual cavity in the absence of coupling) located at $\omega_0 L/c = 19.7896 - 0.5313i$, which has 20 peaks inside the cavity. In the absence of pumping, the poles of the combined system are given by $\omega_- L/c = 19.7404 - 0.1590i$, $\omega_+ L/c = 19.8509 - 0.0741i$. Note that $\text{Re}[\omega_+ - \omega_0] \neq \text{Re}[\omega_0 - \omega_-]$, i.e. a detuning naturally occurs even though the two cavities are identical. Here we assume that the pumps $\gamma_a, \gamma_b$ reduce the absorption (represented by the positive imaginary part of the refractive index $n$) inside the cavity linearly. At threshold $\text{Im}[n]$ becomes negative in at least one cavity, since the gain has to compensate the cavity loss as well. As before, we increase the gain in the coupled cavity system along pumping trajectories similar to those used previously and we find similar qualitative behaviors for the poles of the scattering matrix $\omega_\pm$ for different maximum pump power. In particular, one such trajectory with modes having $\max|\text{Im}\{n\}| = 0.034$ is shown in Fig. 5. Clearly self-termination of the $V_+$ mode occurs at $\gamma_b / \gamma_{\max} \approx 0.105$.

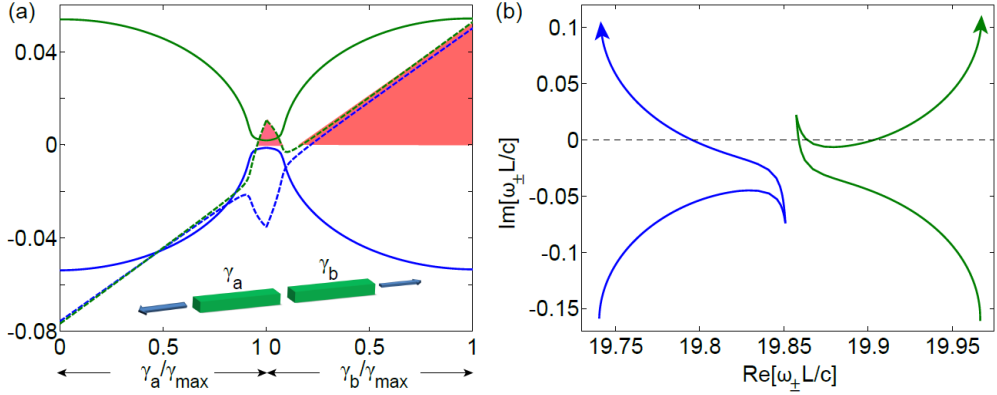

Fig. 5 (a) Ab-initio calculation of the real (solid) and imaginary (dashed) part of $(\omega_\pm - \omega_{avg})L/c$ (green/blue) for two coupled one-dimensional cavities. The red regions show where the $V_+$ mode is lasing. Inset: Schematic of the coupled system. Arrows indicate the laser radiation. (b) Corresponding trajectories of $\omega_\pm$ in the complex plane. The arrows indicate the direction of increasing pump, and two avoided crossings can be identified. The dashed line shows the real axis.

Next we calculate the nonlinear laser intensity using SALT. Here the gain $\gamma$ is represented by the density of the inversion in the lasing medium $D_0$, scaled by its nature unit $d_c = \hbar \gamma_\perp / 4\pi g^2$ to be dimensionless [14]. Here $\hbar$ is Planck's constant, $g$ is the dipole matrix element of the lasing transition and $\gamma_\perp$ is the transverse relaxation rate of the lasing medium. For a pump value $D_0 << |n|^2$ and a lasing mode close to the atomic transition frequency, the main role of $D_0$ is to linearly reduce the absorption and compensate the cavity loss, as in the calculation of the poles discussed above. The corresponding maximum pump value to that in Fig. 5 is $D_{max} \approx 0.25$, given by $2\mathrm{Re}[n]|\Delta\mathrm{Im}[n]|$. The output field intensity $|\psi|^2$ at the left edge of the photonic molecule is shown in Fig. 6(a), and the laser amplitude is expressed in its nature unit $e_c = \hbar\sqrt{\gamma_\parallel \gamma_\perp}/2g$, where $\gamma_\parallel$ is the longitudinal relaxation rate of the lasing medium [14]. Reasonable agreement with the pole calculation is obtained, with self termination of the $V_+$ mode observed at $D_b/D_{max} \approx 0.091$. In Fig. 6(b) we also show that the point at which the self-termination occurs is a function of the maximum pump value. We note that the cavity absorption here ($\text{Im}[n] = 0.007$) is much lower than that used in Ref. 18 ($\text{Im}[n] = 0.13$). It contributes to the loss rate $\kappa$ together with the cavity decay rate, and this observation highlights that the self-termination only requires a $\kappa$ larger than the inter-cavity coupling $J$, which can still be relatively small.

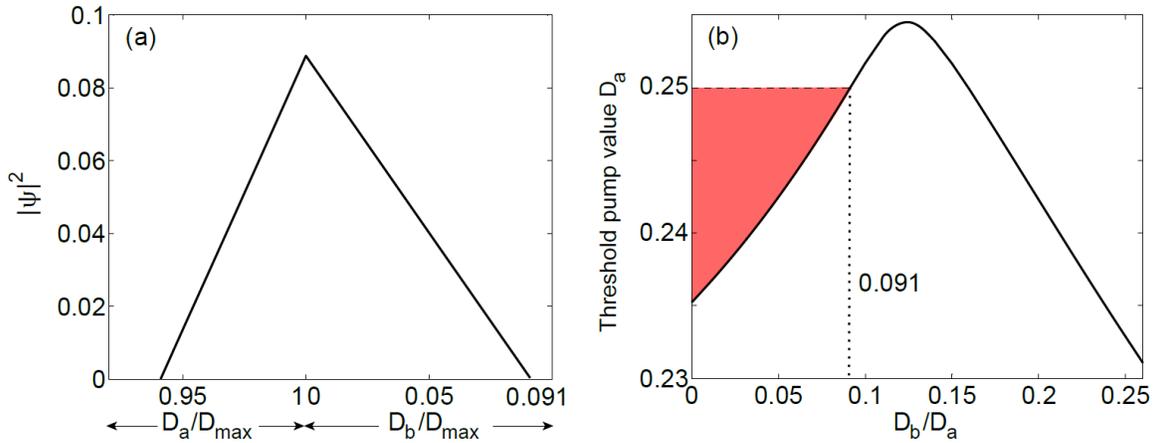

Fig.6 (a) Ab-initio calculation of lasing intensity for the $V_+$ mode at the left edge of the photonic molecule studied in Fig.5. The $V_-$ mode does not lase along this pump trajectory. (b) The threshold pump value of the left cavity as a function of the pump ratio. The horizontal dashed line shows the fixed maximum pump value $D_a = 0.25$ in cavity $a$ as we increase the pump $D_b$ in cavity $b$. The system remains above the threshold (indicated by the red region) until the self-termination point at $D_b/D_a \approx 0.091$, marked by the vertical dotted line.

## 3. Multi-cavity photonic molecules

The discussion in the previous section is by no means pertinent to 2-cavity laser molecules but can be also extended to multi-cavity array. As an example, consider a non-Hermitian system of 4-cavity array described by Hamiltonian:

$$H = \begin{bmatrix} \Omega_o + 3i\chi & \sqrt{3}J & 0 & 0 \\ \sqrt{3}J & \Omega_o + i\chi & 2J & \\ & 2J & \Omega_o - i\chi & \sqrt{3}J \\ & & \sqrt{3}J & \Omega_o - 3i\chi \end{bmatrix}, \qquad (3)$$

where $\Omega_o = \omega_o + i(\gamma - \kappa)$ and the parameterization $\chi$ accounts for the asymmetric pumping. Figure 7(a) shows a schematic of this 4-element photonic molecule where coupling coefficients and asymmetric gain parameters are also indicated.

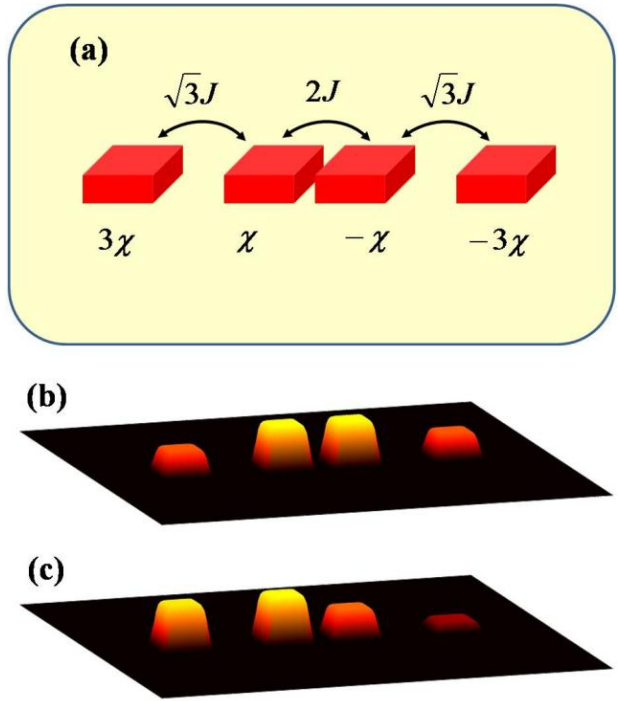

Fig.7 (a) Schematic of the 4-cavity photonic molecule arrangement given by $H$ in Eq.(3). When $J = 1$, $\kappa = 1.5$, $\gamma = 1.5$ and $\chi = 0.5$, we find four symmetric supermodes at lasing threshold point and full SALT equations are necessary to analyze the lasing mode competition. Panel (b) depicts the intensity profile of the eigenmode $\omega_4$ under the above conditions. On the other hand, when $J = 1$, $\kappa = 1.5$, $\gamma = 1.5$ and $\chi = \sqrt{J^2 + \kappa^2/9}$, only $\omega_4$ reaches lasing threshold with an asymmetric intensity distribution as shown in (c). Note that in both (b) and (c), high intensity fields do not reside in the cavity element with the largest gain parameter (leftmost resonator). This peculiar effect does not exist in dimer photonic molecules.

The eigenvalues of this multi-cavity photonic molecule are given by:

$$\omega_n = \omega_o + i(\gamma - \kappa) + (2n-5)\sqrt{J^2 - \chi^2} \quad , \quad n = 1,2,3,4 \qquad (4)$$

Note that this system exhibit a 4<sup>th</sup> order exceptional point at $J = \chi$. Similar to the dimer scenario, we identify two lasing regimes for the lowest threshold mode $\omega_4$. When $J > \chi$ and $\gamma \geq \kappa$, the lasing action of $\omega_4$ will be in the symmetric phase. On the other hand, if the threshold is reached while $J < \chi$, the laser emission profile will be asymmetric. These distinct lasing regimes are depicted in Figs.7 (b) and (c) for the parameters indicated in the figure caption. Note that in both (b) and (c), high intensity fields do not reside in the cavity element with the largest gain parameter (leftmost resonator). This peculiar effect does not exist in dimer photonic molecules.

This analysis reveals that pump-controlled lasing is a universal phenomenon that depends mainly on the exceptional points of the Hamiltonian. We note that the lasing characteristics of the other modes $\omega_{1,2,3}$ can be determined only by taking nonlinear modal interaction into account and thus cannot be captured by the above linear array model. It would be of interest to analyze the effect of these higher order exceptional points on the laser action and we carry this elsewhere.

Next we break the apparent *PT* symmetry of the Hamiltonian (3) by pumping only the left two cavities:

$$H = \begin{bmatrix} \Omega_c + i\gamma_1 & \sqrt{3}J & 0 & 0 \\ \sqrt{3}J & \Omega_c + i\gamma_2 & 2J & 0 \\ 0 & 2J & \Omega_c & \sqrt{3}J \\ 0 & 0 & \sqrt{3}J & \Omega_c \end{bmatrix}, \qquad (5)$$

in which $\Omega_c = \omega_o - i\kappa$. We again increase the pump in cavity 1 to a maximum value while keeping $\gamma_2 = 0$ (trajectory $S_1$), after which we fix $\gamma_1$ and increase $\gamma_2$ (trajectory $S_2$). Similar to the 2-cavity case, we encounter one EP (EP$_1$ in Fig. 8(a)) on $S_1$ for $J=1$ and $\kappa=1.5$, which takes place below the threshold at $\gamma_1 \approx 2.5$ and lasing occurs at $\gamma_1 \approx 2.635$ for a single mode (see Fig. 8(a)). With $\gamma_{max}=2.8$, we encounter another EP (EP$_2$) at $\gamma_2/\gamma_{max} \approx 0.14$ on $S_2$, which occurs again below the threshold, thus preceded by the laser self-termination (at $\gamma_2/\gamma_{max} \approx 0.12$) and followed by revival at $\gamma_2/\gamma_{max} \approx 0.68$. Different from the 2-cavity case, a third EP (EP$_3$) takes place if we further increase the pump in cavity 2 to $\gamma_2/\gamma_{max} \approx 2.51$. This transition is accompanied by a localization transition (see Fig.8(b)) and gives rise to two highly confined modes in cavity 1 and 2, respectively. The latter (mode 2) has the highest net gain as shown in Fig. 8(a). These two modes have very weak overlap and lase simultaneously at frequency $\omega_o$.

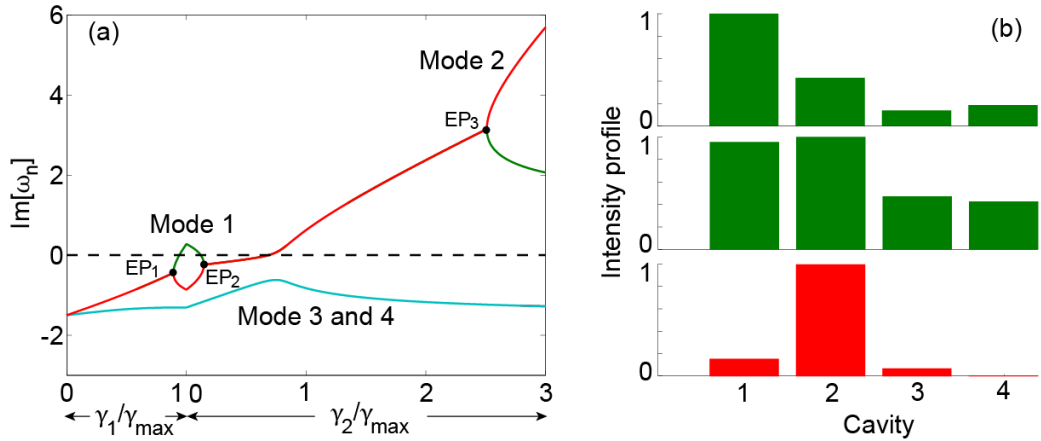

Fig. 8 Pump-induced laser self-termination and localization in a 4-cavity array. (a) The imaginary parts of the four eigenvalues (solid lines). The dashed line shows the real axis. Mode 1 is the only lasing mode before EP$_2$. Its intensity profiles at the onset threshold and self-termination point are shown in the upper and middle panels of (b). Mode 2 is localized in cavity 2 with the highest net gain after EP$_3$. Its intensity profile at $\gamma_2/\gamma_{max}=3$ is shown in the lower panel of (b).

# 4. Conclusion

In conclusion, we have introduced a non-Hermitian dimer model as a means to gain more insight into the behavior of photonic molecule lasers. We have shown that this model predicts laser self-termination in two coupled resonators and provides a clear understanding for the role played by the system's exceptional points. The interplay between all relevant physical parameters (cavity coupling, loss, etc) during this process emerges naturally from the investigated equations. Even more interestingly, we have demonstrated that, for frequency mismatched cavities, operation in the proximity of exceptional points (without actually crossing the square root singularities) can still lead to laser self-termination- an effect that has not been predicted before (fine). In order to confirm our findings, we have compared these latter results with scattering matrix analysis and SALT simulations for 1D two coupled resonators and good agreement were found. We have also used our model to gain an insight into the pump-controlled lasing action. In particular we have shown that laser emission patterns are governed by the locations of the exceptional points of the Hamiltonian. Finally we have extended our results to multi-cavity photonic molecules having higher order exceptional points.

**References**

1. R. K. Chang and A. J. Campillo, *Optical Processes in Microcavities*, Advanced Series in Applied Physics, Vol. 3 (World Scientific, Singapore, 1996).

2. K. J. Vahala, *Optical Microcavities*, Advanced Series in Applied Physics, Vol. 5 (World Scientific, Singapore, 2004).

3. O. Painter et al. Science **284**, 1819 (1999).

4. S. Strauf, K. Hennessy, M. T. Rakher, Y.-S. Choi, A. Badolato, L. C. Andreani, E. L. Hu, P. M. Petroff, and D. Bouwmeester, Phys. Rev. Lett. **96**, 127404 (2006).

5. S. L. McCall, A. F. J. Levi, R. E. Slusher, S. J. Pearton, and R. A. Logan, Appl. Phys. Lett. **60**, 289 (1992).

6. V. S. Ilchenko and A. B. Matsko, IEEE J. Sel. Top. Quantum Electron. **12**, 15 (2006).

7. M. Khajavikhan, A. Simic, M. Katz, J. H. Lee, B. Slutsky, A. Mizrahi, V. Lomakin and Y. Fainman, Nature, **482**, 204 (2012).

8. A. Nakagawa, S. Ishii, and T. Baba, Appl. Phys. Lett. **86**, 041112 (2005).

9. G. Fasching et al. Opt. Express **17**, 20321 (2009).

10. M. Sargent, M. Scully, andW. E. Lamb, Laser Physics (Addison-Wesley, Boston, 1974).

11. H. Haken, Light: Laser Dynamics, Vol. 2 (North-Holland, Amsterdam, 1985).

12. E.I. Smotrova, A.I. Nosich, T.M. Benson, and P. Sewell, IEEE Journal of Selected Topics in Quantum Electronics, **12**, 78 (2006).

13. H. E. Tureci, L. Ge, S. Rotter, and A. D. Stone, Science **320**, 643 (2008).

14. L. Ge, Y. D. Chong, and A. D. Stone, Phys. Rev. A **82**, 063824 (2010).

15. L. Ge, R. J. Tandy, A.D. Stone, and H. E. Tureci, Opt. Express **16**, 16895 (2008).

16. A. Cerjan, Y. D. Chong, L. Ge, and A. D. Stone, Opt. Express **20**, 474 (2012).

17. T. Hisch, M. Liertzer, D. Pogany, F. Mintert, and S. Rotter, Phys. Rev. Lett. **111**, 023902 (2013).

18. M. Liertzer, L. Ge, A. Cerjan, A. D. Stone, H. E. Tureci, and S. Rotter, Phys. Rev. Lett. **108**, 173901 (2012).

19. J. Andreasen, C. Vanneste, L. Ge, and H. Cao, Phys. Rev. A **81**, 043818 (2010).

20. L. Ge, Y. D. Chong, S. Rotter, H. E. Türeci, and A. D. Stone, Phys. Rev. A **84**, 023820, (2011).

21. M. Chitsazi, S. Factor, J. Schindler, H. Ramezani, F. M. Ellis, T. Kottos, arXive:1309.3624.

22. C. M. Bender, and S. Boettcher, Phys. Rev. Lett. **80**, 5243 (1998).

23. A. Guo, G. J. Salamo, D. Duchesne, R. Morandotti, M. Volatier Ravat, V. Aimez, G. A. Siviloglou and D. N. Christodoulides, Phys. Rev. Lett. **103**, 093902 (2009).

24. C. E. Ruter, K.G. Makris, R. El-Ganainy, D.N. Christodoulides, M. Segev, D. Kip, Nature Physics **6**, 192 (2010).

25. Y. D. Chong, L. Ge, and A. D. Stone, Phys. Rev. Lett. **106**, 093902 (2011).

26. P. Ambichl, K. G. Makris, L. Ge, Y. Chong, A. D. Stone, and S. Rotter, Phys. Rev. X **3**, 041030 (2013).